\documentclass[preprint,proceedings]{rmaa}

\SetYear{2002}
\SetConfTitle{Galaxy Evolution: Theory and Observations}

\title{The Link Between Rotation Curve Type and Spiral Arm Structure in Disk
Galaxies}

\author{
  M. S. Seigar\altaffilmark{1}, P. A. James\altaffilmark{2},
I. Puerari\altaffilmark{3}, D. L. Block\altaffilmark{4}}

\altaffiltext{1}{Joint Astronomy Center, 660 N. A'ohoku Place, Hilo, HI 96720,
USA.}
\altaffiltext{2}{Astrophysics Research Institute, Liverpool John Moores
University}
\altaffiltext{3}{INAOE, Puebla, M\'exico.}
\altaffiltext{4}{University of the Witwatersrand, Johannesburg, South
Africa.}

\suppressfulladdresses

\listofauthors{M. S. Seigar, P. A. James, I. Puerari, D. L. Block}
\indexauthor{Seigar, M. S.}
\indexauthor{James, P. A.}
\indexauthor{Puerari, I.}
\indexauthor{Block, D. L.}

\addkeyword{Galaxies: fundamental parameters}
\addkeyword{Galaxies: spiral}
\addkeyword{Galaxies: structure}

\begin{document}
% Typeset article header
\maketitle 

\boldabstract{A new classification scheme for spiral galaxies (Block \& Puerari
1999) has been developed which classifies galaxies on the basis of their
near-IR arm morphology. This `dust-penetrated class' predicts a 
correlation between spiral arm pitch angle and rotation curve
type (or shear rate $A/\omega$, where $A$ is the first Oort constant and 
$\omega$ is rotational velocity - Block et al. 1999; Fuchs 2000). If
such a correlation exists, it would provide a physical basis for this
classification scheme. The Hubble classification breaks down in the
near-IR, where it has been shown that neither pitch angle or $K$
band bulge-to-disk ratio correlate well with Hubble type (Seigar \& James
1998a, b; de Jong 1996). We have therefore investigated the properties of arm
morphology in images of spiral galaxies (e.g. pitch angle) and dynamical
properties from their rotation curves (e.g. shear rate) in order to determine
if there is a physical basis for the dust penetrated class.}

We have observed 8 galaxies at UKIRT with
the imaging camera, UFTI, in the $K$ band to a depth of 21.5 
magnitudes/arcsec$^2$, S/N=3 from August 1st -- 4th, 2001. The sample
was selected from Mathewson et al. (1992). They all have measured H$\alpha$
rotation curves, major axis, $a<1.5$ arcmin, ratio of major-to-minor axes, 
$a/b<3.0$ and declination, $\delta>-25^{\circ}$.

The galaxies were deprojected to a face-on orientation and 
pitch angles were measured using a FFT technique (Schr\"oder et al.
1994). The shear rate is measured from the rotation curves using equation 1.
\begin{equation}
  \label{eq:shear}
  \frac{A}{\omega} = \frac{1}{2}\left(1 - \frac{R}{V} \frac{dV}{dR} \right)
\end{equation}

\begin{figure}[!t]
  \includegraphics[width=5.5cm,angle=-90]{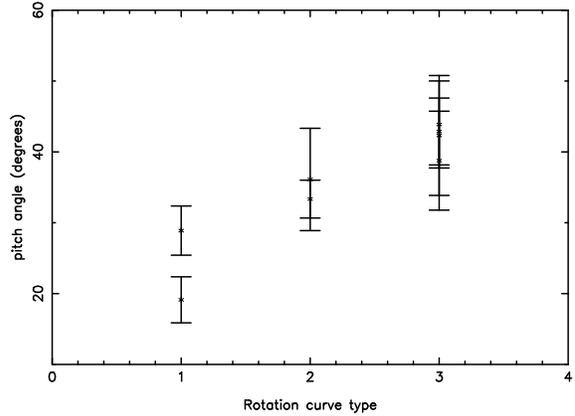}
  \caption{Rotation curve type (1=declining, 2=flat, 3=increasing) versus
    spiral arm pitch angle.}
  \label{fig:rotn}
\end{figure}

\begin{figure}[!t]
  \includegraphics[width=5.5cm,angle=-90]{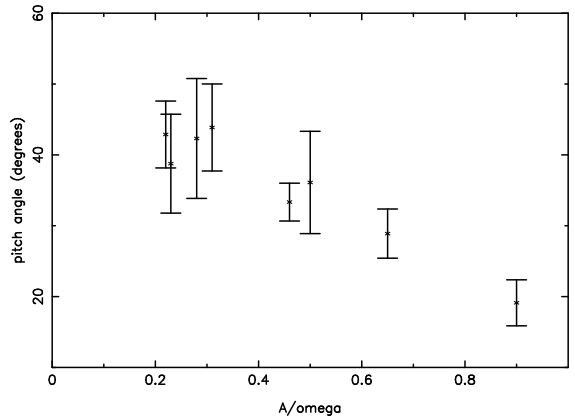}
  \caption{Shear rate ($A/\omega$) versus
    spiral arm pitch angle.}
  \label{fig:rotn}
\end{figure}

Figure 1 shows a correlation between rotation curve type and pitch angle. 
Figure 2 shows a correlation between shear rate and pitch angle. However,
these correlations are only 94\% significant and the errors are $\sim$20\%.
Nevertheless, these are extremely promising initial results, and we now
have more data, going even deeper, which should enable us to strengthen
our conclusions.

%That's all there's room for.


\begin{thebibliography}

\bibitem{B99} Block, D.~L., \& Puerari, I. 1999, A\&A, 342, 627

\bibitem{dJ96} de Jong, R.~S. 1996, A\&A, 313, 45

\bibitem{F00} Fuchs, B. 2000, in {\em Galaxy Dynamics}, eds. F. Combes,
     G.~A. Mamon, ASP Conf. Ser. vol. 197, p 53

\bibitem{M92} Mathewson, D.~S., et al. 1992, ApJS, 81, 413

\bibitem{S94} Schr\"oder, M.~F.~S., et al. 1994, A\&AS, 108, 41

\bibitem{S98a} Seigar, M.~S., \& James, P.~A. 1998a, MNRAS, 299, 672

\bibitem{S98b} Seigar, M.~S., \& James, P.~A. 1998b, MNRAS, 299, 685

\end{thebibliography}
\end{document}